# Revisiting the Majorana Relativistic Theory of Particles with Arbitrary Spin

Luca Nanni*

*E-mail of corresponding author: luca.nanni68@gmail.com

**Abstract**

In 1932 Ettore Majorana published an article proving that relativity allows any value for the spin of a quantum particle and that there is no privilege for the spin ½. The Majorana idea was so innovative for the time that the scientific community understood its importance only towards the end of the thirties. This paper aims to highlight the depth of the scientific thought of Majorana that, well in advance of its time, opened the way for modern particle physics and introduced for the first time the idea of a universal quantum equation, able to explain the behavior of particles with arbitrary spin and of any nature (bradions and tachyons), regardless the value of their speed. It will be analyzed in detail and made explicit all the steps that lead to the physical-mathematical formulation of the Majorana's theory. A part of these steps require basic knowledge of quantum physics but not for this should be regarded as trivial since they show the physical meaning hidden into the structure of the equation. Moreover, the explicit method for the construction of the infinite matrices will be given, by which the infinite components of the wave functions representing the fundamental and excited states of the particle are calculated.

## 1   Introduction

In the early thirties quantum theory was faced with two controversial problems for which it needed a robust explanation to ensure the further development of the still young quantum mechanics: the negative energy solutions of Dirac equation and the interpretation of the continuous spectrum of beta decay. The first was solved in 1932 thanks the positron discovery by C. Anderson [1], the second was only theoretically solved by W. Pauli in 1930 who postulated the existence of the neutrino [2] stating the following: "I have predicted something which shall never be detected experimentally". We know today that for each particle there is a corresponding antiparticle and that the neutrino is a massive particle, even if its repose mass is very small; for both, experimental proofs are now available and are almost irrefutable [3-5]. However, for the physicists of the time none experimental data were available (the technologies used at the time were not yet so advanced to perform sophisticated experiments) and the physical explanations given to the theoretical results, arising from the resolution of the new quantum equations, were somewhat fanciful. In this regard, we recall the Dirac vacuum model seen as an infinite sea of particles with negative energy, whose aim were that to explain the existence of quantum states with negative energy predicted by its famous relativistic equation. In this theoretical model, the antiparticle of the electron was conceived of as a hole in the Dirac's sea. In those years Majorana was working at the institute of Physics of Via Panisperna in the research team of Enrico Fermi. He was very fascinated by the problems above mentioned and, in accordance with his edgy and hypercritical character together with his deep curiosity and astonishing skill in mathematics, did not hesitate to search a robust scientific theory able to explain these problems, criticizing and not accepting a part of the conclusions proposed by Dirac. The questions put by Majorana were: do antiparticles need exist for any spin ½ particles? May spin ½ particles be identical to their antiparticles? Do particles with any arbitrary spin exist? Do particles with arbitrary spin have their own antiparticles? Here the term antiparticle is used, even if in 1932 was yet a meaningless word (at least before the Anderson discovery occurred few months later). Majorana believed that there were not physical reasons that could justify the not-existence of particles with any value of spin: as well as the theory of relativity applied to the quantum mechanics had predicted the electronic spin, so it should do for other hypothetical particles. On the other words, the Dirac equation in principle is correct but neither electron spin nor the existence of its antiparticle was a privilege for the particles with spin ½. Majorana undertaken its work on the basis of this assumption, and with the conviction that the solutions with negative energy were due to an imperfection of the Dirac equation and that could have been eliminated by a mathematical reworking. The entire Majorana work is shown in the bibliographic citation [6]. Due to the publication of the article on an Italian journal of physics (Italian language was unknown for the large part of the scientific community) and due to the completely innovative idea in contrast with the scientific though of the time, the Majorana relativistic theory remained unknown until the end of the thirties when new particles were discovered and the physicists had to formulate a new theory that were a generalization of that of Dirac.

In this paper the Majorana relativistic equation is revisited with the aim to highlight the results that most contributed to the development of the modern quantum theory, focusing the attention whether on its mathematical formalism (Majorana was probably the first physicist to apply the group theory to quantum physics) that on its physical meaning. As will be discussed further, the Majorana equation for particles with arbitrary spin may be seen as the first attempt to find a universal quantum theory able to explain by means of a relativistically invariant equation the physical nature of any particle, regardless the value of its speed. This idea is somewhat revolutionary since we are only in 1932 and the known particles were only the electron and the proton, and the tachyon was not yet even been postulated, being forbidden by the theory of relativity! In the literature there are few articles describing the physical and mathematical foundations on which the Majorana's theory is based, and none of these faces in depth the methodological approach used by the author to formulate his equation [7-9]. In this article we will attempt to fill this gap as fully as possible, even if it means making pedagogical some parts of the discussion.

## 2  The Majorana Equation for Particles with Arbitrary Spin

In this section will be reviewed the entire article of Majorana about the relativistic theory of particles with arbitrary spin [6], explaining the mathematical steps most relevant and commenting critically the assumptions made and the results obtained. In doing so, we will not hesitate in dealing with the latest quantum theories so as to emphasize both the limitations that the innovative aspects, outstanding for the period in which the article was published, of the Majorana theory. As mentioned in the introductory section, Majorana was not satisfied about the Dirac interpretation of the negative energy solutions, although he accepted the mathematical and physical correctness of Dirac equations and was very fascinated by how its relativistic formulation had predicted in a natural way the existence of the spin [10]. It must be remembered that when Majorana formulated its theory, antiparticle was not yet been discovered! The Majorana's purpose was to formulate a generalization of Dirac theory for the electron that was valid for particles with arbitrary spin, that did not include negative energies for not-bound states and it was reduced to the Schrodinger theory for slow motions. Majorana set off by the fact that the relativistically invariant structure of Dirac equation was correct and that it should be kept unchanged for the new generalized equation. This means that Majorana accepted the idea of four-dimensional spinor but he intended to correct the physical meaning of the last two components (i.e. those describing the antiparticle). In other words, the spinor is a state vector describing the single particle and it is formed by four components: the relevance of the first two does not depend on the particle speed while that of the other components is considerable only when the particle speed is close to that of light. Today it's known that the antiparticles are real entities and that, from a mathematical point of view, their spinor components are the complex conjugates of those describing the particle. So, in principle, one of the major assumptions of Majorana is not met. However, his idea led to the hypothesis of a type of particle that is antiparticle of itself and that has been discovery recently [10]. What we thought was a limit of the Majorana's theory led to an important scientific discovery.

The relativistic Dirac equation for the not-bound states of electron may be written as [11]:

$$(H + c\boldsymbol{\alpha} \cdot \hat{p} - \boldsymbol{\beta}mc^2)|\Psi\rangle = 0 \qquad (1)$$

where $\hat{p}$ is the linear momentum operator, H is the relativistic Hamiltonian operator, $\boldsymbol{\alpha}$ is a 3-vector whose components are the $\boldsymbol{\alpha}_k$ 4x4 Dirac matrices and $|\Psi\rangle$ is the spinor. The Dirac matrices $\boldsymbol{\alpha}_k$ and $\boldsymbol{\beta}$ are determined forcing the equation (1) to comply with the energy-momentum relationship $E^2 = m^2c^4 + p^2c^2$. The explicit forms of these matrices are:

$$\boldsymbol{\alpha}_1 = \begin{pmatrix} 0 & 0 & 0 & 1 \\ 0 & 0 & 1 & 0 \\ 0 & 1 & 0 & 0 \\ 1 & 0 & 0 & 0 \end{pmatrix}, \quad \boldsymbol{\alpha}_2 = \begin{pmatrix} 0 & 0 & 0 & -i \\ 0 & 0 & i & 0 \\ 0 & -i & 0 & 0 \\ i & 0 & 0 & 0 \end{pmatrix}, \quad \boldsymbol{\alpha}_3 = \begin{pmatrix} 0 & 0 & 1 & 0 \\ 0 & 0 & 0 & -1 \\ 1 & 0 & 0 & 0 \\ 0 & -1 & 0 & 0 \end{pmatrix}$$

$$\boldsymbol{\beta} = \begin{pmatrix} 1 & 0 & 0 & 0 \\ 0 & 1 & 0 & 0 \\ 0 & 0 & -1 & 0 \\ 0 & 0 & 0 & -1 \end{pmatrix}$$

In this respect, Majorana aims to formulate an equation relativistically invariant without the request to comply with the energy-momentum relationship. Furthermore, Majorana requires that this equation is valid however indeterminate is the particle speed. The latter statement is very important since Majorana has already thought to a theory that can describe the behavior of tachyons.

To avoid the problem of negative energies, Majorana proposed to consider the spinor with infinite many components in such a way that it cannot be split in finite sub-spinors (that is just what is being done to solve the Dirac equation for the free electron). Referring to the relativistically invariant structure of the (1) Majorana applied the variational principle:

$$\delta\langle\Psi|H + c\boldsymbol{\alpha}\cdot\hat{p} - \boldsymbol{\beta}mc^2|\Psi\rangle = \delta(\langle\Psi|H|\Psi\rangle + \langle\Psi|c\boldsymbol{\alpha}\cdot\hat{p}|\Psi\rangle - \langle\Psi|\boldsymbol{\beta}mc^2|\Psi\rangle) = 0 \quad (2)$$

This equation must be valid for particles with arbitrary spin and thus the matrices $\boldsymbol{\alpha}_k$ cannot be equal to those of Dirac but they must coincide with them when the spin is ½. The (2) is Lorentz invariant only if all the components comply this requirement. From the last term of the (2) we get the scalar invariant:

$$\langle\Psi|\boldsymbol{\beta}|\Psi\rangle = \langle\varphi|\varphi\rangle \quad \text{where } |\varphi\rangle = |\Psi\rangle \text{ and } \langle\varphi|\boldsymbol{\beta}^* = \langle\Psi|$$

Replacing the new bra and ket vectors in equation (2) we get:

$$\delta\langle\varphi|\boldsymbol{\gamma}_0 H|\varphi\rangle + \delta\langle\varphi|c\boldsymbol{\gamma}\cdot\hat{p}|\varphi\rangle - mc^2\langle\varphi|\varphi\rangle = 0$$

where we set $\boldsymbol{\gamma}_0 = \boldsymbol{\beta}^{-1}$ and $\boldsymbol{\gamma} = \boldsymbol{\beta}^{-1}\cdot\boldsymbol{\alpha}$. Now, we have to find the form of the $\boldsymbol{\gamma}_k$ matrices, that must assure the invariance of (2), and the transformation law of the spinor $|\varphi\rangle$ respect the elements of the Lorentz group. We are going to face the most complex and cryptic part of the Majorana article; therefore, from now on we will make use of a formalism that differs from that of the original article, but that will clarify the physical concepts.

## 2.1 The Transformation Law of Spinor

Let's start to solve the last point; to avoid any formal complications Majorana made use of unitary infinitesimal transformations given by:

$$\begin{cases} \mathbf{S}_x = \begin{pmatrix} 0 & 0 & 0 & 0 \\ 0 & 0 & 0 & 0 \\ 0 & 0 & 0 & -1 \\ 0 & 0 & 1 & 0 \end{pmatrix}, \quad \mathbf{S}_y = \begin{pmatrix} 0 & 0 & 0 & 0 \\ 0 & 0 & 0 & 1 \\ 0 & 0 & 0 & 0 \\ 0 & -1 & 0 & 0 \end{pmatrix}, \quad \mathbf{S}_z = \begin{pmatrix} 0 & 0 & 0 & 0 \\ 0 & 0 & -1 & 0 \\ 0 & 1 & 0 & 0 \\ 0 & 0 & 0 & 0 \end{pmatrix} \\ \mathbf{T}_x = \begin{pmatrix} 0 & 1 & 0 & 0 \\ 1 & 0 & 0 & 0 \\ 0 & 0 & 0 & 0 \\ 0 & 0 & 0 & 0 \end{pmatrix}, \quad \mathbf{T}_y = \begin{pmatrix} 0 & 0 & 1 & 0 \\ 0 & 0 & 0 & 0 \\ 1 & 0 & 0 & 0 \\ 0 & 0 & 0 & 0 \end{pmatrix}, \quad \mathbf{T}_z = \begin{pmatrix} 0 & 0 & 0 & 1 \\ 0 & 0 & 0 & 0 \\ 0 & 0 & 0 & 0 \\ 1 & 0 & 0 & 0 \end{pmatrix} \end{cases} \quad (3)$$

These are all Lorentz infinitesimal rotations and boosts; any other infinitesimal elements of the Lorentz group can be obtained by means of their linear combination. To get the finite transformations will be sufficient integrate the infinitesimal ones. Using the (3) we construct two hermitian operators $\hat{a}$ and $\hat{b}$ whose Cartesian operatorial elements are:

$$\begin{cases} \hat{a}_x = i\mathbf{S}_x, \quad \hat{a}_y = i\mathbf{S}_y, \quad \hat{a}_z = i\mathbf{S}_z \\ \hat{b}_x = -i\mathbf{T}_x, \quad \hat{b}_y = -i\mathbf{T}_y, \quad \hat{b}_z = -i\mathbf{T}_z \end{cases} \quad (4)$$

To ensure the integrability of infinitesimal Lorentz transformations the operators (4) must satisfy the following commutation relations:

$$[\hat{a}_x, \hat{a}_x] = 0, \quad [\hat{a}_x, \hat{a}_y] = i\hbar \hat{a}_z, \quad [\hat{a}_x, \hat{b}_y] = i\hbar \hat{b}_z, \quad [\hat{a}_x, \hat{b}_z] = -i\hbar \hat{b}_y, \quad \ldots \quad (5)$$

where the ellipsis mean that the other relationships are obtained by means of cyclic permutation of the indices. Relations (5) are the explicit form of the Lie algebra of Lie [11]. Thus, in the Majorana theory the quantum commutation relations arise spontaneously from mathematical requirements, while in the Dirac theory are postulated (canonical quantization) [12]. To better understand the mathematical relations between integrability of infinitesimal transformations and commutators (5) let's consider the following example. Let be $\mathbf{R}_x(\alpha)$ and $R_y(\beta)$ two infinitesimal rotations respectively about x and y axis; their exponential matrix representation is:

$$\mathbf{R}_x(\alpha) = e^{\alpha \mathbf{S}_x} \approx \sum_n \frac{(\mathbf{S}_x)^n}{n!} \quad ; \quad \mathbf{R}_y(\beta) = e^{\beta \mathbf{S}_y} \approx \sum_n \frac{(\mathbf{S}_y)^n}{n!}$$

Performing first a rotation about y axis and then about x axis, the total operator is given by the product of the two infinitesimal rotations:

$$R_x(\alpha) \mathbf{R}_y(\beta) = e^{\alpha \mathbf{S}_x} e^{\beta \mathbf{S}_y} = e^{(\alpha \mathbf{S}_x + \beta \mathbf{S}_y)}$$

Developing the exponential function in Taylor series and recalling that the product between matrices is non-commutative we get:

$$\mathbf{R}_x(\alpha)\mathbf{R}_y(\beta) \approx 1 + (\alpha \mathbf{S}_x + \beta \mathbf{S}_y) + \frac{1}{2}(\alpha \mathbf{S}_x + \beta \mathbf{S}_y)^2 + \frac{1}{2}[\alpha \mathbf{S}_x + \beta \mathbf{S}_y] + \cdots$$

The commutator $[\alpha \mathbf{S}_x + \beta \mathbf{S}_y]$ is added in order to obtain the algebraic form of the square of a binomial that is needed for the integration:

$$(\alpha \mathbf{S}_x + \beta \mathbf{S}_y)^2 + [\alpha \mathbf{S}_x + \beta \mathbf{S}_y] == \alpha^2 \mathbf{S}_x \mathbf{S}_x + \alpha\beta \mathbf{S}_x \mathbf{S}_y + \alpha\beta \mathbf{S}_y \mathbf{S}_x + \beta^2 \mathbf{S}_y \mathbf{S}_y + \alpha\beta \mathbf{S}_x \mathbf{S}_y - \alpha\beta \mathbf{S}_y \mathbf{S}_x =$$

$$= \alpha^2 \mathbf{S}_x \mathbf{S}_x + 2\alpha\beta \mathbf{S}_x \mathbf{S}_y + \beta^2 \mathbf{S}_y \mathbf{S}_y$$

Relations (11) are similar to the typical commutator of the total angular momentum and are satisfied by infinite matrices whose diagonal elements are indexed by j and m, which can be integer or half-integer numbers:

$$j = \frac{1}{2}, \frac{3}{2}, \frac{5}{2}, \ldots \quad \rightarrow \quad m = j, j-1, \ldots, -j$$

$$j = 0, 1, 2, \ldots \quad \rightarrow \quad m = j, j-1, \ldots, -j$$

These numbers are respectively the total angular momentum quantum number, given by the sum of the orbital and spin angular momenta (performed according to the rule of angular momentum composition), and its projection along the z axis. Using the vectorial space of the total angular momentum functions $\{|j, m\rangle\}$, Majorana obtained the non-zero elements of infinite matrices that satisfy the relations (11); they are given by the integrals:

$$\begin{cases}
\begin{cases}
\langle j, m | \hat{a}_x - i\hat{a}_y | j, m+1 \rangle = \hbar \sqrt{(j+m+1)(j-m)} \\
\langle j, m | \hat{a}_x + i\hat{a}_y | j, m-1 \rangle = \hbar \sqrt{(j-m+1)(j+m)} \\
\langle j, m | \hat{a}_z | j+1, m+1 \rangle = \hbar m
\end{cases} \\
\begin{cases}
\langle j, m | \hat{b}_x - i\hat{b}_y | j+1, m+1 \rangle = -\frac{1}{2}\hbar \sqrt{(j+m+1)(j+m+2)} \\
\langle j, m | \hat{b}_x - i\hat{b}_y | j-1, m+1 \rangle = \frac{1}{2}\hbar \sqrt{(j-m)(j-m-1)} \\
\langle j, m | \hat{b}_x + i\hat{b}_y | j+1, m-1 \rangle = \frac{1}{2}\hbar \sqrt{(j-m+1)(j-m+2)}
\end{cases} \\
\begin{cases}
\langle j, m | \hat{b}_x + i\hat{b}_y | j-1, m-1 \rangle = -\frac{1}{2}\hbar \sqrt{(j+m)(j+m-1)} \\
\langle j, m | \hat{b}_z | j+1, m \rangle = \frac{1}{2}\hbar \sqrt{(j+m+1)(j-m+1)} \\
\langle j, m | \hat{b}_z | j-1, m \rangle = -\frac{1}{2}\hbar \sqrt{(j+m)(j-m)}
\end{cases}
\end{cases} \quad (12)$$

We see immediately that there are three integrals involving operators $\hat{a}_i$ (infinitesimal space rotations) and six involving operators $\hat{b}_i$ (infinitesimal boosts). This is due to the fact that operators $\hat{a}_i$ act only on the spatial coordinates while operators $\hat{b}_i$ act on the space-time coordinates. The (12) represent the transitions between quantum states produced by the operators $\hat{a}_x - i\hat{a}_y$, $\hat{a}_x + i\hat{a}_y$ and $\hat{b}_x - i\hat{b}_y$, $\hat{b}_x + i\hat{b}_y$, known as ladder operators and usually labelled by $\hat{a}_\pm$ and $\hat{b}_\pm$, which occur with a non-zero probability. To understand their physical meaning let's consider the product between the operators $\hat{a}_- = \hat{a}_x - i\hat{a}_y$ and $\hat{a}_+ = \hat{a}_x + i\hat{a}_y$:

$$\hat{a}_+\hat{a}_+ = (\hat{a}_x + i\hat{a}_y)(\hat{a}_x - i\hat{a}_y) = \hat{a}_x^2 + \hat{a}_y^2 + i[\hat{a}_x, \hat{a}_y] = \hat{a}_x^2 + \hat{a}_y^2 - \hbar\hat{a}_z \qquad (13)$$

The square operator $\hat{a}^2$ is:

$$\hat{a}^2 = \hat{a}_x^2 + \hat{a}_y^2 + \hat{a}_z^2 \quad \Rightarrow \quad \hat{a}_x^2 + \hat{a}_y^2 = \hat{a}^2 - \hat{a}_z^2$$

Substituting this result in the (13) we get:

$$\hat{a}_+\hat{a}_+ = \hat{a}^2 - \hat{a}_z^2 - \hbar\hat{a}_z$$

For the operator $\hat{a}_-$ we proceed in the same way getting:

$$\hat{a}_-\hat{a}_- = \hat{a}^2 - \hat{a}_z^2 + \hbar\hat{a}_z$$

Let's apply the two operators $\hat{a}_+\hat{a}_+$ and $\hat{a}_-\hat{a}_-$ to the ket $|j, m\rangle$:

$$\hat{a}_+\hat{a}_+|j, m\rangle = \hat{a}^2|j, m\rangle - \hat{a}_z^2|j, m\rangle - \hbar\hat{a}_z|j, m\rangle = [\hbar^2 J(J+1) - \hbar^2 m^2 - \hbar^2 m^2]|j, m\rangle$$

from which we get:

$$\hat{a}_+|j, m\rangle = \hbar\sqrt{(J + m + 1)(J - m)}$$

The same procedure is applied for the operators $\hat{a}_-$ and $\hat{b}_\pm$. Once calculated the matrices $\hat{a}_\pm$ and $\hat{b}_\pm$ we can get the matrix representations of the operators $\hat{a}_x, \hat{a}_y, \hat{a}_z, \hat{b}_x, \hat{b}_y, \hat{b}_z$:

$$\hat{a}_+ + \hat{a}_- = 2\hat{a}_+ \quad \Rightarrow \quad \hat{a}_x = \frac{\hat{a}_+ + \hat{a}_-}{2}$$

and:

$$\hat{a}_+ - \hat{a}_- = 2\hat{a}_+ \quad \Rightarrow \quad \hat{a}_y = -i\frac{\hat{a}_+ + \hat{a}_-}{2}$$

In conclusion, once fixed mass and the spin of a particle we can calculate their spin matrices. From $J_0 = s$ we construct the matrix associated to the quantum number $J_1 = s + 1$ and so forth ad infinitum. Overall, we get a diagonal matrix formed by the sequence of matrices $J_0, J_1, J_2,$ and so on:

$$\begin{pmatrix} J_0 & 0 & 0 & 0 & 0 \\ 0 & J_1 & 0 & 0 & 0 \\ 0 & 0 & J_2 & 0 & 0 \\ 0 & 0 & 0 & J_3 & 0 \\ 0 & 0 & 0 & 0 & \ddots \end{pmatrix}$$

The wave vector of this matrix will have infinite components:

$$|J_0, m_{J_0}; J_1, m_{J_1}; \ldots\rangle = (\Psi_{s,s}, \Psi_{s,s-1}, \ldots, \Psi_{s,-s}, \Psi_{s+1,s+1}, \Psi_{s+1,s}, \ldots, \Psi_{s+1,-(s+1)}, \ldots)$$

So, operator $\hat{a}$ recalls that of the total angular momentum [11] and since it must be a rotation matrix of the Lorentz group we can decide to put it in analogy with the operator $\hat{J}$. We are finally able to give physical meaning to the operators so far seen as abstract objects. The same thing can be done for the operator $\hat{b}$. If we denote by $|\Psi(r,t)\rangle$ a whatever ket and we perform on it a Lorentz transformation limited to the spatial coordinates $|\Psi(r,t)\rangle \to |\Psi'(r',t)\rangle$, then the hermitian operators (10) take the explicit form:

$$\begin{cases} \hat{a}_x = \hbar^{-1}(y\hat{p}_z - z\hat{p}_y) + \frac{1}{2}\sigma_x \\ \hat{a}_y = \hbar^{-1}(x\hat{p}_z - z\hat{p}_x) + \frac{1}{2}\sigma_y \\ \hat{a}_z = \hbar^{-1}(x\hat{p}_y - y\hat{p}_x) + \frac{1}{2}\sigma_z \\ \begin{cases} \hat{b}_x = \hbar^{-1}x\frac{H}{c} + \frac{i}{2}\alpha_x \\ \hat{b}_y = \hbar^{-1}y\frac{H}{c} + \frac{i}{2}\alpha_y \\ \hat{b}_z = \hbar^{-1}z\frac{H}{c} + \frac{i}{2}\alpha_z \end{cases} \end{cases} \quad (14)$$

where $\sigma$ are the spin matrices. The (14) highlights that the three components of the operator $\hat{a}$ are related to the spatial components of the 4-vector angular momentum, while the three components of the operator $\hat{b}$ are related to the time component of the same 4-vector. In other words, the operator $\hat{a}$ contains information related to the spin-orbit coupling while operator $\hat{b}$ is related only to the components of spin. The (14) become clear by comparing them with the tensor of the classical angular momentum:

$$\begin{pmatrix} 0 & -c(tp_x - xE/c^2) & -c(tp_y - yE/c^2) & -c(tp_z - zE/c^2) \\ c(tp_x - xE/c^2) & 0 & L_{xy} & -L_{zx} \\ c(tp_y - yE/c^2) & -L_{xy} & 0 & L_{yz} \\ c(tp_z - zE/c^2) & L_{zx} & -L_{yz} & 0 \end{pmatrix}$$

We proved that integrals (12) are transformations between states with different total angular momentum; those not written are all zero. We have completed the first step to prove the invariance of the equation (8); we must now prove that matrices $\gamma_k$ are the components of a covariant vector transforming according to the metric of Minkowski space.

## 2.2 The Relativistic Transformation of Gamma Matrices

Let's rewrite the Majorana equation as:

$$E_0 \delta\langle\varphi|\gamma_0|\varphi\rangle + cp\delta\langle\varphi|\gamma|\varphi\rangle - mc^2\langle\varphi|\varphi\rangle = 0 \quad (15)$$

The two terms $\langle\varphi|\gamma_0|\varphi\rangle$ and $\langle\varphi|\boldsymbol{\gamma}|\varphi\rangle$ are respectively the charge density and the current density of the particle. To ensure the relativistic invariance of these terms must be satisfied the following commutation relationships:

$$[\gamma_0, \hat{a}_x] = 0 \ , \ [\gamma_0, \hat{b}_x] = i\hbar\gamma_x \ , \ [\gamma_x, \hat{a}_x] = 0 \ , \ [\gamma_x, \hat{b}_y] = i\hbar\gamma_z \ ,$$

$$[\gamma_x, \hat{a}_z] = -i\hbar\gamma_y \ , \qquad [\gamma_x, \hat{b}_x] = i\hbar\gamma_0 \ , \quad \ldots \qquad (16)$$

These relations are very similar to the (11) and this was expected since the matrices $\gamma_k$ are always related to an angular momentum. From the (16) Majorana obtained $\gamma_0 = (J + 1/2)$, while the others are those whose non-zero components satisfy the integrals:

$$\begin{cases}
\langle j, m|\gamma_x - i\gamma_y|j + 1, m + 1\rangle = -\dfrac{i}{2}\sqrt{(j + m + 1)(j + m + 2)} \\
\langle j, m|\gamma_x - i\gamma_y|j - 1, m + 1\rangle = -\dfrac{i}{2}\sqrt{(j - m)(j - m - 1)} \\
\langle j, m|\gamma_x + i\gamma_y|j + 1, m - 1\rangle = \dfrac{i}{2}\sqrt{(j - m + 1)(j - m + 2)} \\
\langle j, m|\gamma_x + i\gamma_y|j - 1, m - 1\rangle = \dfrac{i}{2}\sqrt{(j + m)(j - m - 1)} \\
\langle j, m|\gamma_z|j + 1, m\rangle = \dfrac{i}{2}\sqrt{(j + m + 1)(j - m + 1)} \\
\langle j, m|\gamma_z|j - 1, m\rangle = -\dfrac{i}{2}\sqrt{(j + m)(j - m)}
\end{cases} \qquad (17)$$

As expected the (17) are very similar to the (12). Since $\gamma_0 = \boldsymbol{\beta}^\dagger = \boldsymbol{\beta}^{-1}$ and $\boldsymbol{\beta}^\dagger\boldsymbol{\beta} = \boldsymbol{\beta}^2 = \mathbb{1}$ it follows that:

$$\boldsymbol{\beta} = \left(\frac{1}{j + 1/2}\right)\mathbb{1}$$

At the beginning of the section **2** we set:

$$\langle\Psi|\boldsymbol{\beta}|\Psi\rangle = \langle\varphi|\varphi\rangle$$

We are now able writing the explicit form of the transformation $\Psi \Rightarrow \varphi$ which leads to the unitary form $\langle\varphi|\varphi\rangle$. We completed the work to assure that Majorana equation is relativistically invariant. Their solutions are given by:

$$\langle\varphi_{j,m}| = \langle\Psi_{j,m}|\boldsymbol{\beta} = \langle\Psi_{j,m}|\frac{1}{\sqrt{(j + 1/2)}} \qquad (18)$$

The (18) gives all the components of the spinor once fixed the total angular quantum number. In the frame of reference of the particle $j = 1/2$, and so $\boldsymbol{\beta}$ is unitary and positive unlike that of Dirac.

It's so proved that Majorana theory avoids the problem of negative energies for free particles. For massive particles with zero linear momentum p the diagonal matrix of energies is:

$$E_0 = \boldsymbol{\beta}mc^2 \quad \rightarrow \quad E_0 = \frac{mc^2}{(j + 1/2)} \qquad (19)$$

For half-integer values of j the mass of the particle is $m, m/2, m/3, \ldots$, while for integer values is $2m, 2m/3, 2m/5, \ldots$. Indeed, in this frame of reference the total angular momentum coincides with that of spin. The (19) becomes:

$$E_0 = \frac{mc^2}{(s + 1/2)}$$

If the particle velocity is much lower than that of light, then its mass is:

$$M = \frac{m}{(s + 1/2)}$$

and the spinor components different from zero are those indexed by s and m ($|\Psi_{s,m}\rangle$). In fact, the values of the other components $|\Psi_{s+n,m}\rangle$ (with $n = 1,2,3,\ldots$) are of the order of $(v/c)^n$ and tend to zero as n increases, becoming negligible. To this limit the Majorana equation tends to that of Schrodinger, with the wave function that has only $2s + 1$ non null components (each of which meets separately the Schrodinger equation). The formulation of a theory able of producing solutions with only positive energies is a remarkable achievement for the historical period we are considering: particles with different masses in their frame of reference (where they are at rest) have different intrinsic angular momentum. Majorana was not entirely convinced about this result and to save the physical meaning of his equation he supposed that the only acceptable solution is that of the fundamental state of the particle; all the other states with increasing j must be considered meaningless [8]. However, there are no physical reasons to get rid such solutions that represent transitions from the ground state to exited states with decreasing masses. In fact, if the particle speed approaches that of light, the components of the spinor, depending on $(v/c)^n$, are not more negligible.

As mentioned in the section **2**, Majorana formulates his equation without respecting the energy-momentum relativistic relation. So, for particles with nonzero linear momentum in addition to the states belonging to positive values of the mass, there are others in which energy and linear momentum follow the relation:

$$E^2 = p^2c^2 - m^2c^4$$

that requires $p \geq mc$. These states can be regarded as pertaining to the imaginary value *im* of the mass. In this regard, Majorana introduced in his theory the concept of tachyon, a particle whose velocity exceeds that of light. Whereas the Einstein relativity theory is based on the concept of upper limit of the speed of light, Majorana introduced once again concepts unthinkable for the knowledge of the physicists of the time. The tachyonic behavior of a particle gives further support to the concept of transition between states with different mass that, paradoxically, gradually become the prevailing components of the spinor, being the term $v/c$ greater than 1.

At this point, a question arises: what is the physical interpretation of the Majorana spinor? Well, the most natural interpretation of the spinor is that of a wave vector representing simultaneously all bosons or fermions depending on the intrinsic momentum. Changing the frame of reference by a Lorentz transformation all the spinor components are mixed together without being more separable. If we remain in the frame of reference of the center of mass, the spinor transforms like that of a particle with a given spin. In the Majorana theory fermions and bosons are treated in a completely symmetrical way, as well as the real and imaginary masses.

## 3 Conclusion

Despite the depth and the modernity of the ideas that characterize the Majorana theory, the equation for particles with arbitrary spin remains a subject still little known and used in the field of particle physics. In fact, the physics of Majorana is mainly cited in the scientific literature limited to research on neutrinos or on the Majorana fermions, which refer to his latest article published in 1937 [13]. The goal of this article is to bring to the attention of the modern research on particle physics a theory that could contribute to its further development, overcoming the impasse in which today we are with the modern field theory. This equation, in fact, has been formulated without the obligation to respect the law of conservation of energy or the upper limit of the speed of light, but only in the full respect of the Lorentz invariance. This, perhaps, could be the key to get out of this impasse and give new breath to a branch of physics that is hard to progress further.

## 4 References

[1] C. Aderson, H.L. Anderson, (1983). Unraleving the Particle Content of Cosmic Rays, in L.M. Brown and L. Hoddeson (ed.) The Birth of Particle Physics, Cambridge Univ. Press.

[2] G.Marx, (1995). Nucl.Phys.B (Proc.Suppl.) 38, 518.

[3] L. M. Brown, (1978). "The Idea of the Neutrino". , 31 (9): 23–28 - doi:10.1063/1.2995181.

[4] F. Close, (2010). Neutrino, Oxford University Press, ISBN 978-0-19-957459-9.

[5] S. Weinberg, (1995). The quantum theory of fields, Volume 1: Foundations. Cambridge University Press, ISBN 0-521-55001-7.

[6] E. Majorana, (1932) Teoria Relativistica di Particelle con Momento Intrinseco Arbitrario, Il Nuovo Cimento, Vol. 9, pp. 335-344. English translation by C. A. Orzalesi in Technical Report no. 792, (1968), University of Maryland.

[7] D.M. Fradkin, (1966). Comments on a Paper by Majorana Concerning Elementary Particles, American Journal of Physics, Volume 34, Issue 4, 314-318.

[8] R. Casalbuono, Majorana and the Infinite Component Wave Equations, arXiv.org (23 Oct. 2006), http://arxiv.org/pdf/hep-th/0610252.pdf.

[9] Laura Deleichi and Marta Greselin, (2015). About Majorana's Unpublished Manuscripts or Relativistic Quantum Theory of Particles of Any Spin, Universal Journal of Physics and Applications 9(3), 168-171.

[10] S. Nadj-Perge et al., (2014). Science, Vol. 346, n° 6209, pp. 602-607 - DOI: 10.1126/science.1259327.

[11] F.W Warner, (1983). Foundations of Differentiable Manifolds and Lie Groups, Springer – New York.

[11] D.J. Griffiths, (1995). Introduction to Quantum Mechanics. Prentice Hall.

[12] Dirac, P. A. M.. (1925). "The Fundamental Equations of Quantum Mechanics". Proceedings of the Royal Society A: Mathematical, Physical and Engineering Sciences 109 (752) – 642.

[13] E. Majorana, (1937). Teoria Simmetrica dell'Elettrone e del Positrone, Il Nuovo Cimento, Vol. 14, pp. 171-184.